\documentclass[amsmath,amssymb,onecolumn, a4paper,14pt]{revtex4}
\usepackage{amsmath}
\usepackage{amssymb}
\usepackage{amsmath,amsthm,amssymb,amscd}
\usepackage{latexsym}
\usepackage{indentfirst}
\usepackage{subfigure}
\usepackage{graphicx}
\usepackage{extarrows}
\usepackage{bm}
\usepackage{multirow}
\usepackage[colorlinks,linkcolor=blue,hyperindex,dvipdfm]{hyperref}

\date{\today}

\begin{document}

\title{Theoretical analysis of transcription process with polymerase stalling
}

\author{Jingwei Li and Yunxin Zhang} \email[Email: ]{xyz@fudan.edu.cn}
\affiliation{Laboratory of Mathematics for Nonlinear Science, Shanghai Key Laboratory for Contemporary Applied Mathematics, Centre for Computational Systems Biology, School of Mathematical Sciences, Fudan University, Shanghai 200433, China. }

\begin{abstract}
Experimental evidences show that in gene transcription, RNA polymerase has the possibility to be stalled at certain position of the transcription template. This may be due to the template damage, or protein barriers. Once stalled, polymerase may backtrack along the template to the previous nucleotide to wait for the repair of the damaged site, or simply bypass the barrier or damaged site and consequently synthesize an incorrect messenger RNA, or degrade and detach from the template. Thus, the {\it effective} transcription rate (the rate to synthesize correct product mRNA) and the transcription {\it effectiveness} (the ratio of the {\it effective} transcription rate to the {\it effective} transcription initiation rate) are both influenced by polymerase stalling events. So far, no theoretical model has been given to discuss the gene transcription process including polymerase stalling. In this study, based on the totally asymmetric simple exclusion process (TASEP), the transcription process including polymerase stalling is analyzed theoretically. The dependence of the {\it effective} transcription rate, {\it effective} transcription initiation rate, and transcription {\it effectiveness} on the transcription initiation rate, termination rate, as well as the backtracking rate, bypass rate, and detachment (degradation) rate when stalling, will be discussed in detail. The results showed that backtracking restart after polymerase stalling is an ideal mechanism to increase both the {\it effective} transcription rate and transcription {\it effectiveness}. Without backtracking, detachment of stalled polymerase can also help to increase the {\it effective} transcription rate and transcription {\it effectiveness}. Generally, the increase of bypass rate of the stalled polymerase will lead to the decrease of the {\it effective} transcription rate and transcription {\it effectiveness}. But when both detachment rate and backtracking rate of the stalled polymerase vanish, the {\it effective} transcription rate may also be increased by the bypass mechanism.
\end{abstract}

%\keywords{TASEP, gene transcription, polymerase stalling}
%\pacs{87.16.aj, 87.15.rp, 05.70.-a, 87.16.Uv, 05.40.Jc}

\maketitle

\section{Introduction}
Replication, transcription, and translation are three basic processes in cells. Before cell division, a cell replicates its DNA with the help of DNA polymerase. Using DNA as a template, messenger RNA (mRNA) is synthesized by RNA polymerase (RNAP) during the so-called transcription process. Then using mRNA as a template, peptide chain is synthesized by ribosome during the translation process, and proteins are then obtained by the folding of peptide chains. Roughly speaking, each of the three processes includes three subprocesses, initiation, elongation, and termination. %, since a template consists of a series of sites (nucleic acids for DNA and codons for mRNA).
The product is synthesized by polymerase during its forward motion along   template in the elongation process.

In the field of theoretical studies, transcription process is usually described by the totally asymmetric simple exclusion processes (TASEP), see \cite{Krug1991,Derrida1993,Schutz1993,Kolomeisky1998,Zhang20113,Zhang20131}. In which RNAP is regarded as a point particle, and the template DNA is regarded as a one-dimensional lattice with lattice sites corresponding to the nucleotides in DNA. The transcription initiation corresponds to the binding of RNAP to the first lattice site, where the first site can be regarded as a combination of the promoter and the transcription start site. The transcription termination corresponds to the leaving of particle from the last site of the lattice. The elongation of transcription is described by the forward hopping of particle in the main body of the lattice. The totally asymmetric exclusion means that the polymerase at site $i$ can only hopping forward to site $i+1$ provided the site $i+1$ is not occupied. In TASEP, the forward hopping rates of particle at any site $i$ of the lattice are always assumed to be the same and simply normalized to be 1. It implies that RNAP will move along DNA template with constant speed until the termination site. However, several experimental observations found that the regular elongation procedure may be interrupted, with RNAP stalled at certain nucleotide. The stalling of RNAP may be caused by several reasons. Structural aberrations of the template can trigger a stalling of polymerase \cite{Azvolinsky2009,Dalgaard2001,Brewer1988}. Polymerase may also be stalled from the depletion of building blocks NTP \cite{Gruber2000}, or from the template damage \cite{Nouspikel2002,Casper2002,Cha2002,Andreea2014,Damsma2007,Walmacq2012}. Meanwhile, the damage or incorrect assembling of polymerase itself may also lead to stalling \cite{Edenberg2014}.

In both prokaryotic and eukaryotic cells, there are several mechanisms which are usually employed by polymerase to solve the stalling problem. If the stalling is caused by template damage, polymerase may backtrack along the template to the previous site and wait for the repair of the damaged site \cite{Giannattasio2004,Ciccia2012,Weston2012,Yuan2012,Vermeulen2013,Lagerwerf2011,Deaconescu2013,Higgins1976}. The synthesis of mRNA is able to restart after the repair. Alternatively, the stalled polymerase may simply bypass the damaged site and continue the transcription process from the downstream site, and finally end the transcription at the termination site with an incorrect product  \cite{Stelter2003,Watanabe2004,Ciccia2012,Weston2012,Yuan2012,Walmacq2012}. Meanwhile, if a prolonged stalling occurs, the polymerase may be degraded as a mechanism of last resort \cite{Wilson2013}. By the way, in translation process, recent experiments have also found that the template (mRNA) can degrade when the translocation of ribosome is stalled \cite{Shoemaker2012,Doma2006,Becker2011,Shoemaker2010}.

The polymerase stalling as well as the possible mechanisms employed by the stalled polymerase will affect the overall transcription rate and efficiency, and consequently have influence on the strength of gene expression. Thus, the related properties of transcription are not only determined by the initiation rate and termination rate as implied in the usual TASEP model, but also influenced by the polymerase stalling and corresponding mechanisms used to overcome the stalling problem. Although there are various kinds of generalizations of TASEP model, no one can be used directly to describe the gene transcription process with polymerase stalling.
%\cite{Krug1991,Schutz2003,Lipowsky2006,Raguin2013,Popkov2013,Bressloff2013,Zhang20113,Zhang20131,Gupta2014}.

\begin{figure}
  \centering
  \includegraphics[width=12cm]{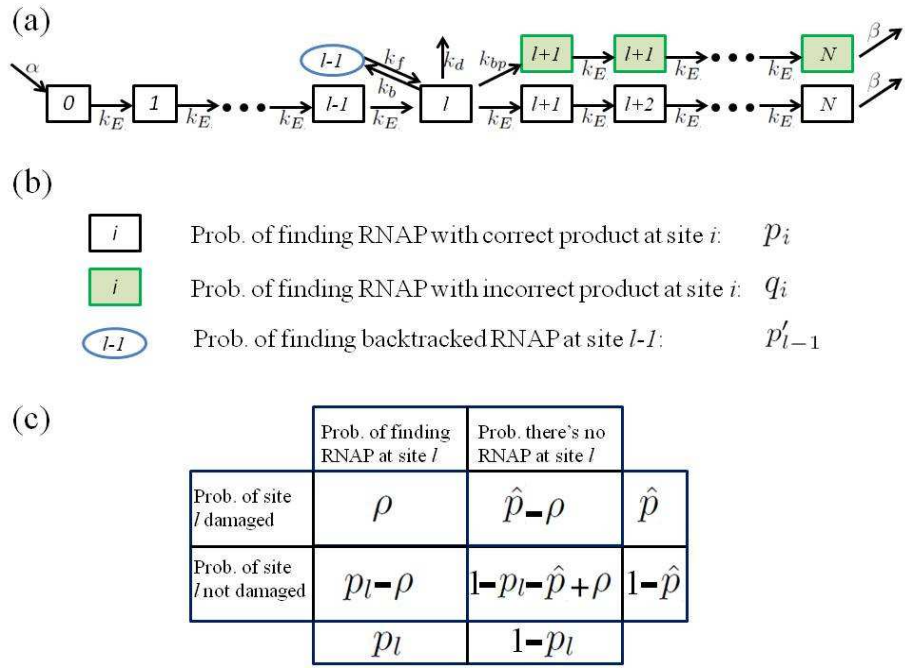}\\
  \caption{ {\bf (a)} Modified TASEP model to describe gene transcription process with possible polymerase stalling at site $l$. Transcription starts with polymerase binding to the first site 0 (with the rate denoted by $\alpha$), and terminated at the last site $N$ with rate denoted by $\beta$. At site $l$, the forward translocation of polymerase may be stalled. There are three mechanisms for a stalled polymerase to leave the damaged site $l$, backtracking to site $l-1$ with rate $k_b$, degrading and detaching from the template with rate $k_d$, or just bypassing the site $l$ with rate $k_{bp}$ and continuing its translocation along the template (but the mRNA synthesized by it is incorrect and will degrade soon). During transcription elongation period, the forward stepping rate of polymerase is denoted by $k_E$, which is assumed to be the same throughout the transcription, for polymerases in whatever states (correct or incorrect). {\bf (b)} Notations for probabilities of finding polymerase at corresponding states, with correctly transcribed mRNA ($p_i$), incorrectly transcribed mRNA ($q_i$), and backtracked polymerase at site $l-1$ ($p'_{l-1}$). {\bf (c)} Notations for probabilities related to site $l$, which may be damaged with or without polymerase binding.}\label{model}
\end{figure}

In this study, a modified TASEP model is presented to describe the gene transcription process including polymerase stalling. For simplicity, this study assumes that there is only one nucleotide in the transcription template at which polymerase may be stalled, and the position of this nucleotide is unchanged for any polymerase. This nucleotide may be damaged, or bound by protein complexes, or there is one special secondary structure around it. The stalled polymerase may backtrack along template to the previous binding site to wait for the repair of the damaged site (or clearance of the barrier), or simply bypass this nucleotide and synthesize an incorrect mRNA product, or degrade and detach from the transcription template, see Fig. \ref{model}. The numerical calculations of our modified TASEP model show that, the {\it effective} transcription rate may be enlarged by increasing the backtracking rate, detachment rate, and bypass rate of the stalled polymerase. Even the transcription {\it effectiveness} may be increased with the backtracking and detachment rates. Generally, backtracking is one ideal mechanism to solve the polymerase stalling problem. Without backtracking, detachment and bypass are also good mechanisms to increase the {\it effective} transcription rate.

This study is organized as follows. The modified TASEP model describing the transcription process including polymerase stalling will be presented in the next section, and then the results obtained by this model will be given in Section III. Finally, concluding remarks will be presented in the last section.

\section{Modified TASEP model for gene transcription with polymerase stalling}
The model used in this study can be regarded as a modification of the usual TASEP, which is schematically depicted in Fig. \ref{model}. Where the length of gene is assumed to be $N+1$, and each lattice site stands for one nucleotide or nucleotide group (which means this model is obtained by coarse grain).  The transcription begins with RNA polymerase binding to lattice site 0 (corresponding to the promoter upstream the gene) with rate $\alpha$, which depends on the concentration of free polymerase in environment, the binding rate of transcription factors, and the nucleotide sequence of promoter \cite{Li2014}.  The transcription is ended by polymerase leaving from lattice site $N$, with corresponding rate denoted by $\beta$. This study assumes that only lattice site $l$ may be damaged (or occupied by a protein complex). The rate constant that site $l$ becomes damaged is denoted by $k_{\hat{p}d}$. If the damaged site $l$ is not occupied by a polymerase, then it can be repaired with rate $k_{\hat{p}r}$. See the following Eq. (\ref{dhatp}) for the dynamics of probability that site $l$ is damaged. If the site $l$ is damaged, polymerase on it may bypass it directly (with no transcription) and continue its forward translocation along template, and finally leave from the stop site $N$, but the product (i.e. mRNA) synthesized by it will be incorrect and will degrade soon. In this study, the probability that there is a polymerase with a {\it correct} semi-finished product at site $i$ (for $0\le i\le N$) is denoted by $p_i$, and the probability that there is a polymerase with an {\it incorrect} semi-finished product at site $i$ (for $l+1\le i\le N$) is denoted by $q_i$. A polymerase at site $i$ (for $i\ne l$) will move to site $i+1$ with rate $k_E$ provided site $i+1$ is unoccupied.

If the site $l$ is damaged, polymerase at this site will be stalled. Experiments found that there are three possible mechanisms for the stalled polymerase to leave the damaged site $l$. {\bf (1)} The polymerase may backtrack to the previous site $l-1$ with rate $k_b$ provided the site $l-1$ is not occupied. After the repair of site $l$, the backtracked polymerase will return to site $l$ with rate $k_f$. The probability of finding a backtracked polymerase at site $l-1$ is denoted by $p'_{l-1}$. See the following Eq. (\ref{dpp}) for the dynamics of probability $p'_{l-1}$. {\bf (2)} The polymerase may bypass the damaged site $l$ with rate $k_{bp}$, and continue its transcription from the downstream site $l+1$. {\bf (3)} The polymerase may degrade with rate $k_d$ and detach from the transcription template.
Note that the genetic information coded in damaged site $l$ cannot be transcribed. Therefore, the mRNA synthesized by a polymerase which has bypassed the damaged site $l$ is nonfunctional and will degrade soon.   %the stalled polymerase at damaged site $l$ cannot move to the next site $l+1$. Bypass means that if the site $l$ is damaged, then the polymerase at site $l-1$ steps forward to site $l+1$ directly instead of entering into the damaged site $l$.
Meanwhile, the damaged site $l$ cannot be repaired if there is a polymerase binding on it. Thus, if $k_b=0$, $k_{bp}=0$, and $k_d=0$, the polymerase will be stalled at the damaged site $l$ forever. So bypass, degradation, and backtracking are three important mechanisms for cells to continue the transcription process. Otherwise, the template will be totally blocked and become useless, and should be degraded.

This study assumes that each site $i$ can only be occupied by one polymerase. If there is one backtracked polymerase at site $l-1$, the site $l$ will be unoccupied. This is because that the backtracked polymerase at site $l-1$ is from site $l$. In the following, the probability of finding a polymerase at damaged site $l$ is denoted by $\rho$. For the model depicted in Fig. \ref{model}, the probabilities $p_i$ %q_i, p'_{l-1}, \rho$
are governed by the following equations
%Secondly, we further consider the probability $\rho$ that there is a polymerase in site $l$, and site $l$ is damaged. $\rho$ is involved because there may exist potential correlations between these two accidents. Only considering the marginal distributions may distort the truth. In fact, if $k_b=0$, $k_{bp}=0$, and $k_d=0$, a polymerase will be stalling in site $l$ forever in the steady state. However, the model without the consideration of $\rho$ lead to a bistable state as showed in Fig. \ref{Tmix}{\bf (a)}, which is obviously wrong.
%We illustrate the equations in our models as follows.
\begin{equation}\label{dp}
\begin{array}{lcl}
dp_0/dt&=&\alpha(1-p_0)-k_Ep_0(1-p_1),\\
dp_i/dt&=&k_Ep_{i-1}(1-p_i)-k_Ep_i(1-p_{i+1}), \qquad\textrm{for } 1\le i\le l-3, \\
dp_{l-2}/dt&=&k_Ep_{l-3}(1-p_{l-2})-k_Ep_{l-2}(1-p_{l-1}-p'_{l-1}),\\
dp_{l-1}/dt&=&k_Ep_{l-2}(1-p_{l-1}-p'_{l-1})-k_Ep_{l-1}(1-p_l), \\
dp_{l}/dt&=&k_Ep_{l-1}(1-p_l)-k_E(p_l-\rho)(1-p_{l+1}-q_{l+1})+k_f(1-\hat{p})p'_{l-1}\\
    &&-k_b\rho(1-p_{l-1}-p'_{l-1})-k_{bp}\rho(1-p_{l+1}-q_{l+1})-k_d\rho, \\
dp_{l+1}/dt&=& k_E(p_l-\rho)(1-p_{l+1}-q_{l+1})-k_Ep_{l+1}(1-p_{l+2}-q_{l+2}), \\
dp_i/dt&=&k_Ep_{i-1}(1-p_{i}-q_{i})-k_Ep_{i}(1-p_{i+1}-q_{i+1}), \qquad\textrm{for }  l+2\le i\le N-1, \\
dp_{N}/dt&=&k_Ep_{N-1}(1-p_N-q_N)-\beta p_N.
\end{array}
\end{equation}
Where the equations for $0\le i\le l-3$ and $l+2\le i\le N$ can be obtained similarly as in usual TASEP model. The total probability of finding polymerase at site $l-1$ is $p_{l-1}+p'_{l-1}$, where $p_{l-1}$ is the probability of polymerase which comes from site $l-2$, and $p'_{l-1}$ is the probability of polymerase which is backtracked to site $l-1$ from the damaged site $l$.  The probability flux from site $l-2$ to site $l-1$, which is related to the governing equations of probabilities $p_{l-2}$ and $p_{l-1}$, is $k_E p_{l-2}(1-p_{l-1}-p'_{l-1})$. In the governing equation of probability $p_l$, the first term is the flux from site $l-1$ to site $l$. The second term is the flux from undamaged site $l$ to site $l+1$, where $p_l-\rho$ is the probability that there is a polymerase at site $l$ and the site $l$ is not damaged. The third term is the return flux from site $l-1$ to site $l$ of the backtracked polymerase, where $1-\hat{p}$ is the probabilty that the damaged site $l$ has been repaired. The forth term is the backtraking flux. The fifth term is the bypass flux, and the final term is the detachment flux. The governing equation for probability $p_{l+1}$ can be obtained similarly.

Meanwhile, the probabilities $q_i$ satisfy (see Fig. \ref{model}{\bf (b)} for the meanings of probabilities $q_i$)
\begin{equation}\label{dq}
\begin{array}{lcl}
dq_{l+1}/dt&=&k_{bp}\rho(1-p_{l+1}-q_{l+1})-k_Eq_{l+1}(1-p_{l+2}-q_{l+2}),\\
dq_{i}/dt&=&k_Eq_{i-1}(1-p_i-q_i)-k_Eq_i(1-p_{i+1}-q_{i+1}),\qquad\textrm{for } l+2 \le i \le N-1, \\
dq_{N}/dt&=&k_Eq_{N-1}(1-p_N-q_N)-\beta q_N.
\end{array}
\end{equation}
Where the probability $\hat{p}$ that site $l$ is damaged satisfies
\begin{equation}\label{dhatp}
d\hat{p}/dt=k_{\hat{p}d}(1-\hat{p})-k_{\hat{p}r}(\hat{p}-\rho).
\end{equation}
In which, the second term is from the assumption that only the unoccupied site $l$ can be repaired. The probability $\rho$ that there is a polymerase at the damaged site $l$ can be obtained as follows
\begin{equation}\label{drho}
d\rho/dt=k_{\hat{p}d}(p_l-\rho)+k_Ep_{l-1}(\hat{p}-\rho)-k_{bp}\rho(1-q_{l+1}-p_{l+1})-k_d\rho-k_b\rho(1-p'_{l-1}-p_{l-1}).
\end{equation}
Where the first term is the flux of probability that the occupied site $l$ becomes damaged. The second term is the flux of probability that a polymerase translocates from site $l-1$ to the unoccupied but damaged site $l$. The last three terms are bypass flux, detachment flux, and backtracking flux, respectively.
Finally, the probability $p'_{l-1}$ that there is a backtracked polymerase at site $l-1$ satisfies
\begin{equation}\label{dpp}
dp'_{l-1}/dt=k_{b}\rho(1-p_{l-1}-p'_{l-1})-k_{f}(1-\hat{p})p'_{l-1}.
\end{equation}
Where the first term is the backtracking probability flux of the stalled polymerase from the damaged site $l$ to its upstream site $l-1$, and the second term is the return probability flux of the backtracked polymerase.
For convenience, meanings of probabilities $p_l, \hat{p}, \rho$ are displayed in Fig. \ref{model}{\bf (c)}. The total probability of finding a polymerase at site $i$, no matter whether it is with a correctly synthesized mRNA or an incorrect mRNA, is denoted by $P_i$, or mathematically,
\begin{equation}\label{Pi}
P_i=\left\{
      \begin{array}{cc}
        p_i, & {\rm for}\ 0\le i\le l-2\ {\rm or}\ i=l, \\
        p_{l-1}+p'_{l-1}, & {\rm for}\ i=l-1, \\
        p_i+q_i, & {\rm for}\ l+1\le i\le N.
      \end{array}
    \right.
\end{equation}

\section{results}
\begin{figure}
  \centering
  \includegraphics[width=12cm]{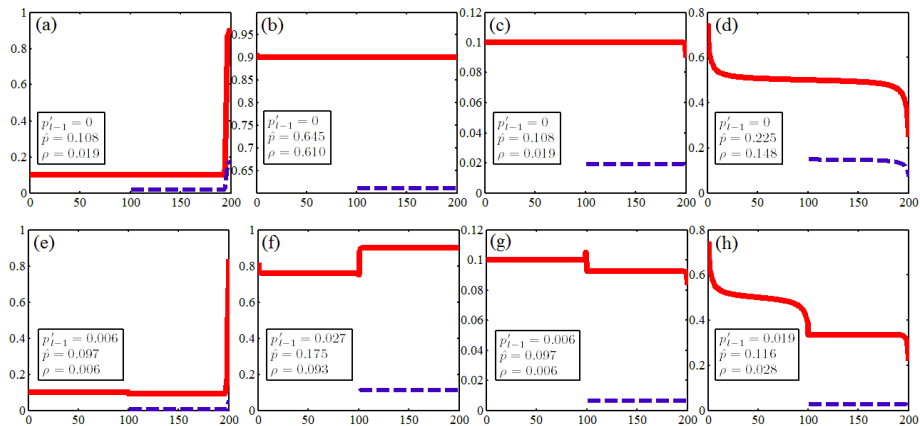}\\
  \caption{Typical examples of probabilities $P_i$ (solid lines), $q_i$ (dashed lines), $p'_{l-1}$, $\hat{p}$, and $\rho$ (given in legends) along the gene, which are obtained from Eqs. (\ref{dp}-\ref{Pi}) with gene length $N=100$. {\bf (a-d)} are for simplified cases where detachment and backtracking of polymerase from damaged site $l$ are not allowed, i.e. $k_b=k_d=0$. While {\bf (e-h)} are for cases with nonzero detachment rate and backtracking rate. Except the values of $k_b$ and $k_d$, other parameter values used in {\bf (a-d)} are the same as the ones used in {\bf (e-h)} respectively, see Table \ref{Tmixtable}. {\bf (a,c,e,g)} are examples of the low probability density case with probability less than 0.5, {\bf (b,f)} are examples of high density case with probability larger than 0.5, and {\bf (d,h)} are examples of maximal flux case. The sharp decrease of probability $P_i$ after site $l=100$ is due to the polymerase detachment from site $l$. For the meanings of probability notations, see Fig. \ref{model}}\label{Tmix}
\end{figure}

All results of this study are based on the steady state solution of Eqs. (\ref{dp}-\ref{dpp}), which are obtained by numerical calculations performed in software MATLAB.
To illustrate the properties of gene transcription with possible stalling of polymerase at a given position, typical examples of related probabilities, obtained by the modified TASEP model, are plotted in Fig. \ref{Tmix}. Where Figs. \ref{Tmix}{\bf (a-d)} are for the cases where $k_b=k_d=0$, i.e., the backtracking rate to the upstream site $l-1$ and detachment rate from site $l$ for stalled polymerase at site $l$ vanish. For these special cases, the total probability $P_i$ of finding polymerase at site $i$ may have three different phases, low density phase (Figs. \ref{Tmix}{\bf (a,c)}), high density phase (Figs. \ref{Tmix}{\bf (b)}), and maximal flux phase (Figs. \ref{Tmix}{\bf (d)}). Where the probability flux is defined by $J_i=K_EP_i(1-P_{i})$, which reaches its maximal value 1/4 when $P_i\equiv1/2$. Meanwhile, boundary layers may exist at one or both of the two boundaries, $i=0$ and $i=N$. These properties are similar as the ones of usual TASEP models, and the probability and corresponding flux are fully determined by the transcription initiation rate $\alpha$ and the transcription termination rate $\beta$, see \cite{Schutz1993,Derrida19931}.
For general cases with nonzero values of backtracking rate $k_b$ and detachment rate $k_d$, the plots in Figs. \ref{Tmix}{\bf (e-h)} show that, the probability $P_i$ has a sharp change at the site $l$. For the sake of comparison, except $k_b$ and $k_d$, other parameter values used in {\bf (e)-(h)} are the same as the ones used in {\bf (a)-(d)} respectively, see Table \ref{Tmixtable}.
Since polymerase at damaged site $l$ may degrade and detach from the transcription template, different with the ones plotted in Figs. \ref{Tmix}{\bf (a-d)}, the probability flux $J_i=K_EP_i(1-P_{i})$ for the general cases is not conversed along the template. Meanwhile, since polymerase can only detach or backtrack from the damaged site $l$, the flux $J_i$ is conversed both in the region between site $i=0$ and site $i=l-1$ and in the region between site $i=l+1$ and site $i=N$. Because of the detachment, the probability flux will be reduced after site $l$. It means that for the cases of low density phase and maximal flux phase, the probability $P_i$ will be reduced after site $l$ (see Figs. \ref{Tmix}{\bf (e,g,h)}), while for the cases of high density phase, $P_i$ will be increased (see Figs. \ref{Tmix}{\bf (f)}). Due to the backtracking of polymerase from damaged site $l$, the total probability of finding polymerase at site $l-1$, $P_{l-1}$, may be higher than those of other sites, see Figs. \ref{Tmix}{\bf (g)}.
\begin{table}
\center
\begin{tabular}{|c|ccccccccccc|}
  \hline
  figures  &\  $N$\  &\  $k_E$\  &\  $l$\  &\  $\alpha$\  &\  $\beta$\  &\  $k_b$\  &\  $k_{bp}$\  &\  $k_d$\  &\  $k_f$\  &\  $k_{\hat{p}d}$\  &\  $k_{\hat{p}r}$\  \\\hline
  Fig. \ref{Tmix}{\bf (a)}  & 200 & 1 & 100 & 0.1 & 0.1 & 0 & 1 & 0 & 1 & 0.1 & 1 \\\hline
  Fig. \ref{Tmix}{\bf (b)}  & 200 & 1 & 100 & 1 & 0.1 & 0 & 1 & 0 & 1 & 0.1 & 1 \\\hline
  Fig. \ref{Tmix}{\bf (c)}  & 200 & 1 & 100 & 0.1 & 1 & 0 & 1 & 0 & 1 & 0.1 & 1 \\\hline
  Fig. \ref{Tmix}{\bf (d)}  & 200 & 1 & 100 & 1 & 1 & 0 & 1 & 0 & 1 & 0.1 & 1 \\\hline
  Fig. \ref{Tmix}{\bf (e)}  & 200 & 1 & 100 & 0.1 & 0.1 & 1 & 1 & 1 & 1 & 0.1 & 1 \\\hline
  Fig. \ref{Tmix}{\bf (f)}  & 200 & 1 & 100 & 1 & 0.1 & 1 & 1 & 1 & 1 & 0.1 & 1 \\\hline
  Fig. \ref{Tmix}{\bf (g)}  & 200 & 1 & 100 & 0.1 & 1 & 1 & 1 & 1 & 1 & 0.1 & 1 \\\hline
  Fig. \ref{Tmix}{\bf (h)}  & 200 & 1 & 100 & 1 & 1 & 1 & 1 & 1 & 1 & 0.1 & 1 \\
  \hline
\end{tabular}
\caption{Parameter values used in the calculations of Fig. \ref{Tmix}.
}\label{Tmixtable}
\end{table}

The plots in Figs. \ref{Tmix}{\bf (a-d)} imply that, without polymerase detachment, the total transcription rate may not be reduced by the site damage. But the {\it effective} (or {\it correct}) transcription rate, i.e., the rate to synthesize correct mRNA, will be reduced. Given the nonzero bypass rate $k_{bp}$, some products are incorrect and will degrade rapidly. With additional detachment of polymerase from the damaged site $l$, for general cases, the {\it effective} transcription rate is less than the {\it effective} transcription initial rate, see Figs. \ref{Tmix}{\bf (e-h)}. In this study, the {\it effective} transcription initial rate is defined as $\alpha_{eff}:=\alpha(1-p_0)$, the {\it effective} transcription rate is defined as $\beta_{eff}:=\beta p_N$, and the {\it bypass} transcription rate is defined as $\beta_{bp}:=\beta q_N$. In the following, the parameter dependent properties of $\alpha_{eff}$, $\beta_{eff}$, $\beta_{bp}$, and the ratio $r:=\beta_{eff}/\alpha_{eff}$ will be discussed in detail. The ratio $r$ is one reasonable index to describe the {\it effectiveness} of transcription including polymerase stalling.

\begin{figure}
  \centering
  \includegraphics[width=12cm]{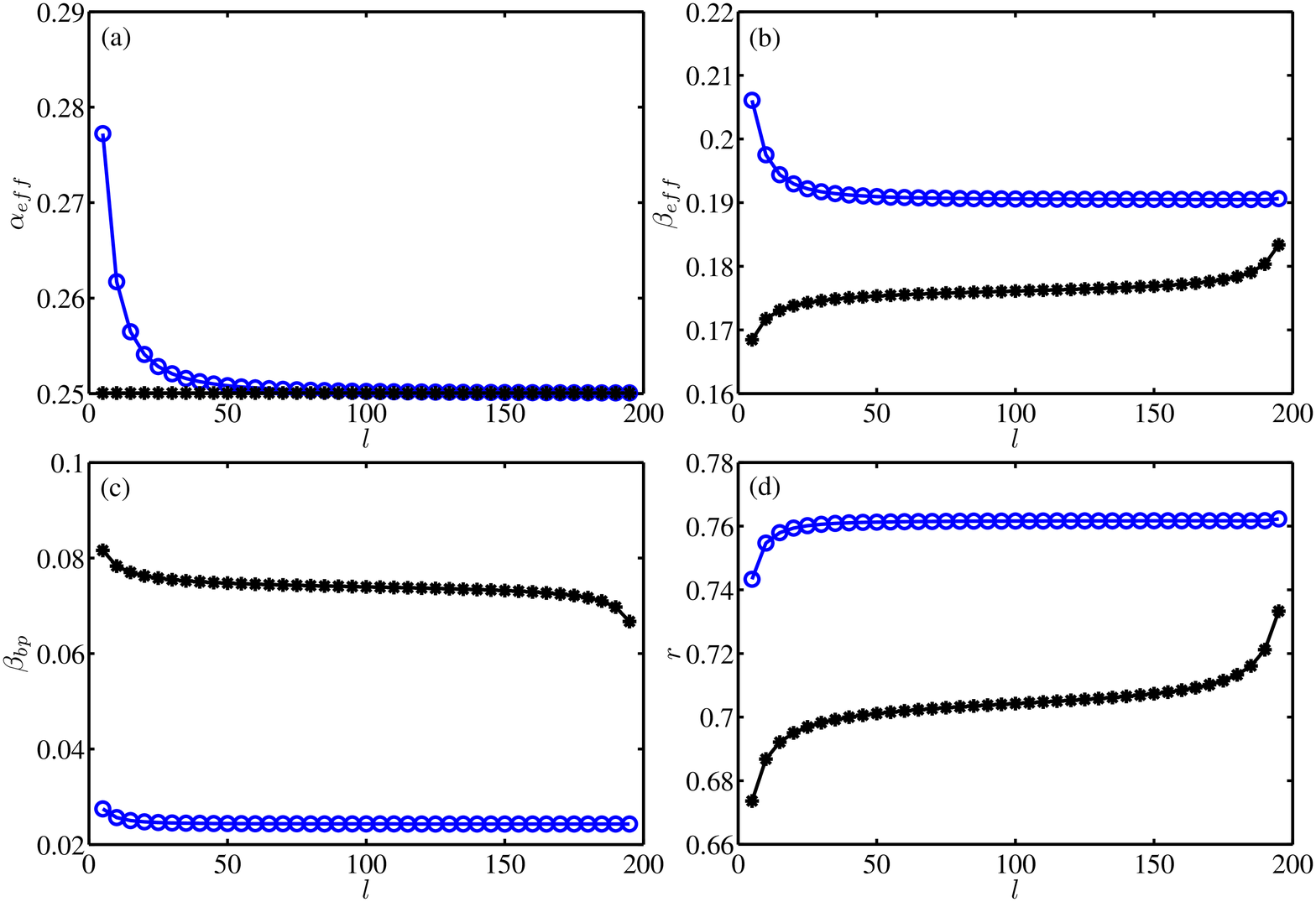}\\
  \caption{The {\it effective} transcription initiation rate $\alpha_{eff}:=\alpha(1-p_0)$ {\bf (a)}, {\it effective} (or correct) transcription rate $\beta_{eff}:=\beta p_N$ {\bf (b)}, {\it bypass} transcription rate $\beta_{bp}:=\beta q_N$ {\bf (c)}, and the transcription {\it effectiveness} $r:=\beta_{eff}/\alpha_{eff}$ {\bf (d)}, as functions of the position $l$ of damaged site, which changes from 10 to 190 with an increment 10. For other parameter values, see Table \ref{paratable}. The only difference between the two lines in each figure is that, for the lines with marker \lq$\ast$', the detachment rate $k_d$ is equal to zero, while for the lines with marker \lq$\circ$', the detachment rate $k_d$ is nonzero. }\label{ldense}
\end{figure}

The plots in Fig. \ref{ldense} show that with nonzero detachment rate $k_d$, all the {\it effective} rates $\alpha_{eff}$, $\beta_{eff}$, and $\beta_{bp}$ decrease with the position $l$ of damaged site, but the transcription {\it effectiveness} $r$ increases with $l$ (see the lines in Fig. \ref{ldense} with marker \lq\lq $\circ$"). Given that polymerase can only detach from the damaged site $l$, if the location $l$ of damaged nucleotide is far from the initiation site $0$, then the polymerase density between sites $0$ and $l$ will be high and consequently the polymerase current along gene will be low. Thus, the {\it effective} transcription initiation rate $\alpha_{eff}$ is decreased with $l$. Except from the damaged site $l$, polymerase can not detach from the transcription template, therefore low {\it effective} initiation rate will lead to low transcription rate. Thus, {\it effective} transcription rate $\beta_{eff}$ and {\it bypass} transcription rate $\beta_{bp}$ also decrease with the damaged position $l$. The increase of transcription {\it effectiveness} $r$ with damaged position $l$ implies that large values of damaged position $l$ will be beneficial for cells to increase the transcription efficiency and save energy molecules. On the other hand, for non-detachment cases, i.e., $k_d=0$ (see the lines in Fig. \ref{ldense} with marker \lq\lq $\ast$"), the {\it effective} initiation rate $\alpha_{eff}$ is independent of damaged position $l$, but the {\it effective} transcription rate $\beta_{eff}$ increases with $l$. Thus the transcription {\it effectiveness} $r=\beta_{eff}/\alpha_{eff}$ also increases with damaged position $l$. Therefore, for any case (with or without detachment of polymerase from the damaged site), large values of damaged position $l$ will help to increase the transcription efficiency. The plots in Fig. \ref{ldense} also show that except for the cases where the damaged site of template is close to the transcription start site or termination site, $\alpha_{eff}$, $\beta_{eff}$, $\beta_{bp}$, and $r$ are not sensitive to the damaged position $l$.
\begin{table}
\center
\begin{tabular}{|c|c|ccccccccccc|}
  \hline
  figures & label & $N$  &  $k_E$ & $l$ & $\alpha$ & $\beta$ & $k_b$ & $k_{bp}$ & $k_d$ & $k_f$ & $k_{\hat{p}d}$ & $k_{\hat{p}r}$ \\\hline
Figs. \ref{ldense}{\bf (a,b,c,d)} & $\circ$   & 200  & 1 &   & 1 & 1 & 0 & 1 & 1 & 1 & 0.1 & 1 \\
    & $\ast$   & 200  & 1 &   & 1 & 1 & 0 & 1 & 0 & 1 & 0.1 & 1 \\
    \hline
    \hline
Figs. \ref{alphadense}{\bf (a,b,c,d)} & $\circ$   & 200  & 1 & 100 &   & 0.1 & 0 & 1 & 1 & 1 & 0.1 & 1 \\
    & $\ast$   & 200  & 1 & 100 &   & 1 & 0 & 1 & 1 & 1 & 0.1 & 1 \\
    & -   & 200  & 1 & 100 &   & 0.1 & 0 & 1 & 0 & 1 & 0.1 & 1 \\
  \hline
  \hline
Figs. \ref{betadense}{\bf (a,b,c,d)} & $\circ$   & 200  & 1 & 100 & 1 &   & 1 & 1 & 0 & 1 & 0.1 & 1 \\
    & $\ast$    & 200  & 1 & 100 & 1 &   & 0 & 1 & 1 & 1 & 0.1 & 1 \\
  \hline
  \hline
Figs. \ref{kbdense}{\bf (a,b,c,d)} & $\circ$   & 200  & 1 & 100 & 1 & 1 &   & 1 & 0 & 1 & 0.1 & 1 \\
    \hline
    \hline
Figs. \ref{kbpdense}{\bf (a,b,c,d)} & $\circ$   & 200  & 1 & 100 & 1 & 1 & 0 &   & 0 & 1 & 0.1 & 1 \\
    & $\ast$   & 200  & 1 & 100 & 1 & 1 & 1 &   & 0 & 1 & 0.1 & 1 \\
    & -   & 200  & 1 & 100 & 1 & 0.1 & 0 &   & 1 & 1 & 0.1 & 1  \\
    \hline
    \hline
Figs. \ref{kddense}{\bf (a,b,c,d)} & $\circ$   & 200  & 1 & 100 & 1 & 0.1 & 0 & 1 &   & 1 & 0.1 & 1 \\
    & $\ast$   & 200  & 1 & 100 & 1 & 0.1 & 1 & 1 &   & 1 & 0.1 & 1 \\
    & -    & 200  & 1 & 100 & 1 & 1 & 1 & 1 &   & 1 & 0.1 & 1 \\
    \hline
    \hline
Figs. \ref{kfdense}{\bf (a,b,c,d)} & $\circ$   & 200  & 1 & 100 & 1 & 1 & 1 & 1 & 0 &   & 0.1 & 1 \\
    & $\ast$   & 200  & 1 & 100 & 1 & 0.1 & 1 & 1 & 1 &   & 0.1 & 1 \\
    \hline
    \hline
Figs. \ref{khatpddense}{\bf (a,b,c,d)} & $\circ$   & 200  & 1 & 100 & 1 & 1 & 1 & 1 & 0 & 1 &   & 1 \\
    & $\ast$   & 200  & 1 & 100 & 1 & 1 & 1 & 0 & 1 & 1 &   & 1 \\
    & -    & 200  & 1 & 100 & 1 & 0.1 & 0 & 1 & 1 & 1 &   & 1 \\
    \hline
%  \hline
\end{tabular}
\caption{Parameter values used in the calculations of Figs. \ref{ldense}-\ref{khatpddense}.
%\ref{alphadense}, \ref{betadense}, \ref{kbdense}, \ref{kbpdense}, \ref{kddense}, \ref{kfdense}, and \ref{khatpddense}.
}\label{paratable}
\end{table}

\begin{figure}
  \centering
  \includegraphics[width=12cm]{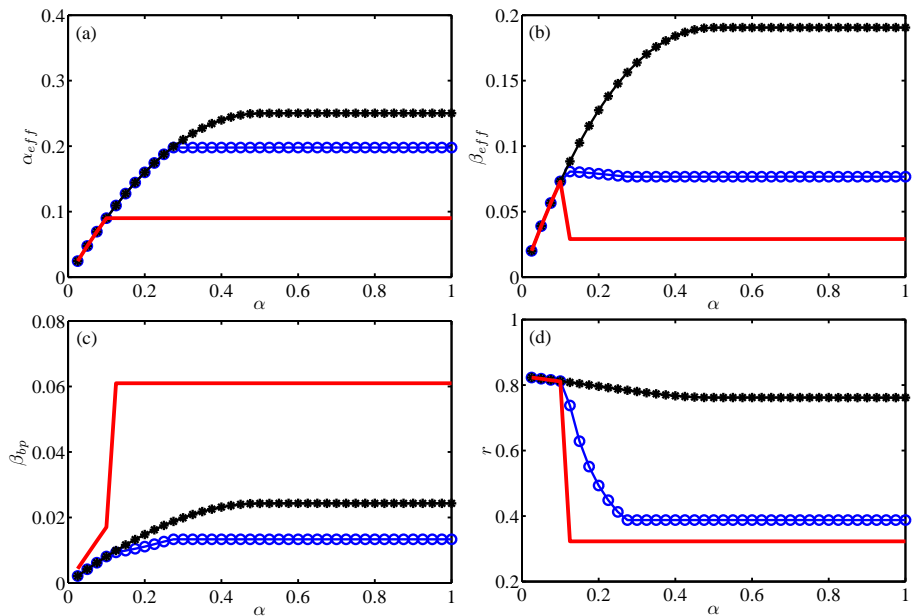}\\
  \caption{The {\it effective} transcription initiation rate $\alpha_{eff}$ {\bf (a)}, {\it effective} transcription rate $\beta_{eff}$ {\bf (b)}, {\it bypass} transcription rate $\beta_{bp}$ {\bf (c)}, and the {\it effectiveness} $r:=\beta_{eff}/\alpha_{eff}$ {\bf (d)}, as functions of the transcription initiation rate $\alpha$. In each figure, three typical examples are plotted. %, with markers \lq$\circ$' and \lq$\ast$', respectively.
  In calculations, the initiation rate $\alpha$ changes from 0.1 to 1 with an increment 0.1, and other parameter values are listed in Table \ref{paratable}. In contrast to the lines with marker \lq$\circ$', the thick solid lines are obtained with zero detachment rate, i.e., $k_d=0$, while the lines with marker \lq$\ast$' are obtained with larger termination rate $\beta$.}\label{alphadense}
\end{figure}

Fig. \ref{alphadense}{\bf (a)} shows that the {\it effective} transcription initiation rate $\alpha_{eff}$ increases with the initiation rate $\alpha$, and tends to approach one limit value. In the calculations of Fig. \ref{alphadense}{\bf (a)}, the line with marker \lq\lq $\ast$" is obtained with large termination rate $\beta$ and nonzero detachment rate $k_d$, the line with marker \lq\lq $\circ$" is obtained with small termination rate $\beta$ and nonzero detachment rate $k_d$, and the thick solid line is obtained with small termination rate $\beta$ and zero detachment rate. Thus, the plots in Fig. \ref{alphadense}{\bf (a)} also imply that the initiation rate limit of the {\it effective} rate $\alpha_{eff}$ increases with the termination rate $\beta$ and detachment rate $k_d$. For large initiation rate $\alpha$, the {\it effective} transcription rate $\beta_{eff}$ also has one limit value, which increases with termination rate $\beta$ and detachment rate $k_d$, see Fig. \ref{alphadense}{\bf (b)}. But, different with the {\it effective} initiation rate $\alpha_{eff}$, $\beta_{eff}$ may not change monotonically with initiation rate $\alpha$. The plots in Fig. \ref{alphadense}{\bf (c)} show that the {\it bypass} transcription rate $\beta_{bp}$ increases with the initiation rate $\alpha$, and tends to one limit value when $\alpha$ is large enough. %. But different with $\alpha_{eff}$,
The limit value of $\beta_{bp}$ increases with termination rate $\beta$ but decreases with detachment rate $k_d$. This is because, for large values of detachment rate $k_d$, polymerase will have less chances to reach the stop site of the template. Finally, the transcription {\it effectiveness} $r$ decreases with initiation rate $\alpha$, and its limit value increases with both the termination rate $\beta$ and detachment rate $k_d$, see Fig. \ref{alphadense}{\bf (d)}.

\begin{figure}
  \centering
  \includegraphics[width=12cm]{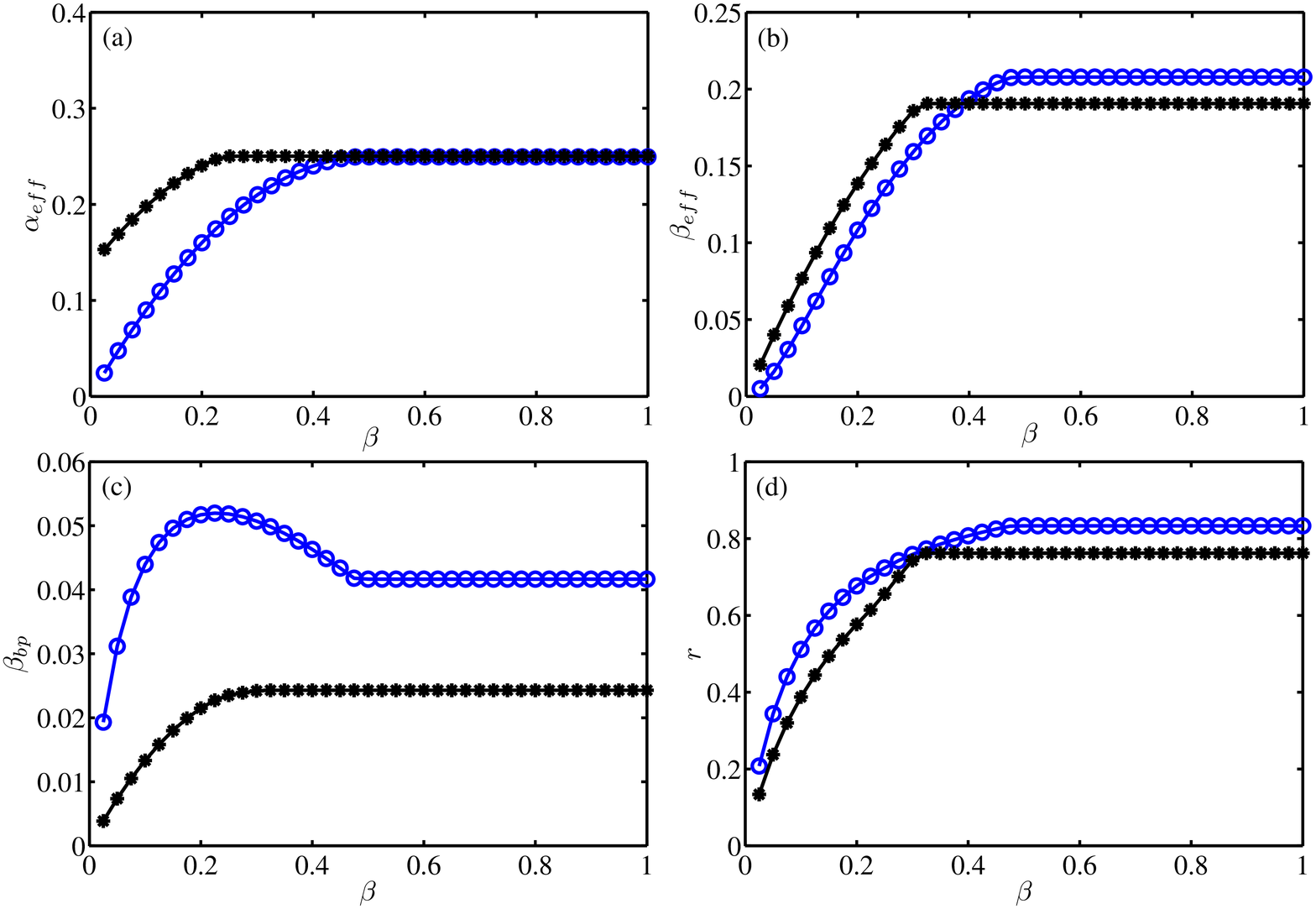}\\
  \caption{The {\it effective} transcription initiation rate $\alpha_{eff}$ {\bf (a)}, {\it effective} transcription rate $\beta_{eff}$ {\bf (b)}, {\it bypass} transcription rate $\beta_{bp}$ {\bf (c)}, and  transcription {\it effectiveness} $r$ {\bf (d)}, as functions of the termination rate $\beta$. The lines with marker \lq$\circ$' are obtained with zero detachment rate $k_d$ but nonzero backtracking rate $k_b$, while the lines with marker \lq$\ast$' are obtained with nonzero detachment rate $k_d$ but zero backtracking rate $k_b$. In calculations, the termination rate $\beta$ changes from 0.1 to 1 with an increment 0.1, for other parameter values, see Table \ref{paratable}.}\label{betadense}
\end{figure}

Except the {\it bypass} transcription rate $\beta_{bp}$, both of the two rates $\alpha_{eff}$ and $\beta_{eff}$, and the {\it effectiveness} $r$, increase monotonically with the termination rate $\beta$, and tend to approach corresponding limit values for large $\beta$, see Fig. \ref{betadense}. The backtracking of stalled polymerase at damaged site $l$ can help to raise the transcription {\it effectiveness} $r$ (see the line with marker \lq\lq$\circ$" in Fig. \ref{betadense}{\bf (d)}). For high termination rate $\beta$, $\beta>0.5$, backtracking also helps to raise the {\it effective} transcription rate $\beta_{eff}$ (see the plots in Fig. \ref{betadense}{\bf (b)}). Therefore, high termination rate and backtracking rate are beneficial to getting high {\it effective} transcription rate and to increasing the transcription {\it effectiveness}. With backtracking but no detachment, the stalled polymerase at damaged site $l$ will have additional chance to continue its transcription.

\begin{figure}
  \centering
  \includegraphics[width=12cm]{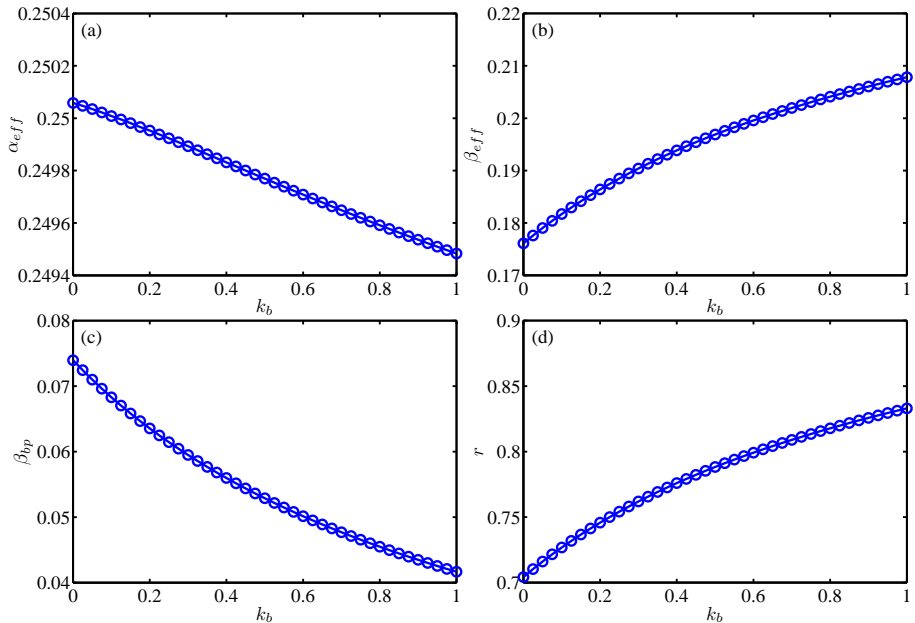}\\
  \caption{The {\it effective} transcription initiation rate $\alpha_{eff}$ {\bf (a)}, {\it effective} transcription rate $\beta_{eff}$ {\bf (b)}, {\it bypass} transcription rate $\beta_{bp}$ {\bf (c)}, transcription {\it effectiveness} $r$ {\bf (d)}, as functions of the backtracking rate $k_b$. In calculations, $k_b$ changes from 0 to 1 with an increment 0.1. The detachment rate $k_d$ is set to zero, i.e. stalled polymerases at damaged site $l$ will not detach from the template. Other parameter values used in calculations are listed in Table \ref{paratable}.}\label{kbdense}
\end{figure}

Without detachment, i.e., $k_d=0$, there are only two mechanisms for the stalled polymerase to leave damaged site $l$, backtracking to site $l-1$ and waiting for the repair of site $l$ or bypassing the damaged site $l$ and continuing its transcription from site $l+1$. With the increase of backtracking rate $k_b$, the translocation of polymerase along the template will be slowed down. Thus, the {\it effective} transcription initiation rate $\alpha_{eff}$ decreases with backtracking rate $k_b$, see Fig. \ref{kbdense}{\bf (a)}. Given that there are only two mechanisms for the stalled polymerase to leave damaged site $l$, the increase of backtracking rate will lead to the decrease of the probability of bypass. This implies that, the {\it bypass} transcription rate decreases with backtracking rate $k_b$, see Fig. \ref{kbdense}{\bf (c)}. Finally, the plots in Figs \ref{kbdense}{\bf (b,d)} show that, both the {\it effective} transcription rate $\beta_{eff}$ and the transcription {\it effectiveness} $r$ increase with backtracking rate $k_b$. Thus, backtracking is one of the ideal mechanisms for cells to solve the stalling problem.

\begin{figure}
  \centering
  \includegraphics[width=12cm]{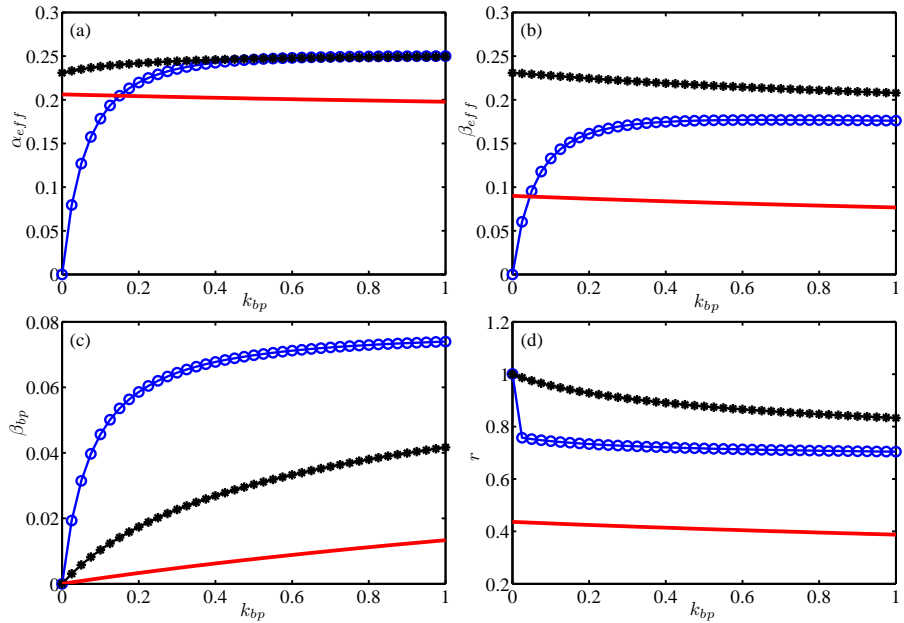}\\
  \caption{The {\it effective} transcription initiation rate $\alpha_{eff}$ {\bf (a)}, {\it effective} transcription rate $\beta_{eff}$ {\bf (b)}, {\it bypass} transcription rate $\beta_{bp}$ {\bf (c)}, and the transcription {\it effectiveness} $r$ {\bf (d)}, as functions of the bypass rate $k_{bp}$. In each figure, three typical examples are plotted. Where the lines with marker \lq$\ast$' are obtained with $k_d=0$ and $k_b\ne0$, the thick solid lines are obtained with $k_d\ne0$ and $k_b=0$, and the lines with marker \lq$\circ$' are obtained with $k_d=k_b=0$. The bypass rate $k_{bp}$ changes from 0 to 1 with an increment 0.1. The values of other parameters are listed in Table \ref{paratable}.}\label{kbpdense}
\end{figure}

If the stalled polymerase can only continue its translocation by the bypass mechanism, i.e. bypass the damaged site $l$ and continue its transcription from site $l+1$, and cannot backtrack to site $l+1$ or detach from the template, then the {\it effective} transcription initiation rate $\alpha_{eff}$, the {\it bypass} transcription rate $\beta_{bp}$, and the {\it effective} transcription rate $\beta_{eff}$ will all increase with the bypass rate $b_{bp}$, see Fig. \ref{kbpdense}{\bf (a-c)}. The increase of rate $\beta_{eff}$ with bypass rate $b_{bp}$ is because, with large values of $k_{bp}$, the polymerase with correctly synthesized mRNA will have less possibility to be blocked during its transcription process. However, the plots in Fig. \ref{kbpdense}{\bf (d)} indicate that transcription {\it effectiveness} $r$ decreases with bypass rate $k_{bp}$. Besides the bypass mechanism, if the stalled polymerase can also backtrack to the previous site $l-1$ to wait for the repair of the damaged site $l$, then the rates $\alpha_{eff}, \beta_{eff}$, and {\it effectiveness} $r$ will be increased (see the lines with markers \lq\lq$\circ$" and \lq\lq$\ast$" in Figs. \ref{kbpdense}{\bf (a,b,d)}). The lines plotted in Figs. \ref{kbpdense}{\bf (c)} with markers \lq\lq$\circ$" and \lq\lq$\ast$" show that, with additional backtracking mechanism, i.e., $k_{b}\ne0$, the {\it bypass} transcription rate $\beta_{bp}$ will be reduced. Meanwhile, all the solid lines and the lines with marker \lq\lq$\circ$" in Figs. \ref{kbpdense}{\bf (a-d)} show that with additional detachment mechanism, i.e., $k_d\ne0$, all the {\it effective} rates $\alpha_{eff}, \beta_{eff}$, and $\beta_{bp}$, and the transcription {\it effectiveness} $r$ will be reduced.
 This implies that the detachment of stalled polymerase may not be one good mechanism for cells to solve the transcription stalling problem, and to increase their transcription rate and efficiency. The plots in Fig \ref{kbpdense} also show that, for the special cases with either nonzero detachment rate or nonzero backtracking rate, the {\it effective} rates $\alpha_{eff}, \beta_{eff}, \beta_{bp}$ only change slightly with the bypass rate $k_{bp}$.

\begin{figure}
  \centering
  \includegraphics[width=12cm]{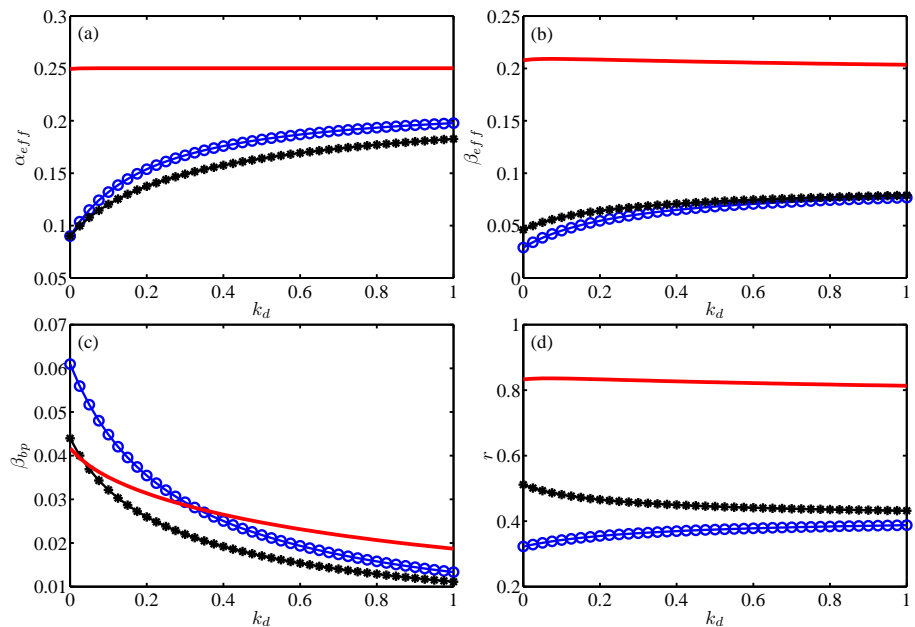}\\
  \caption{The {\it effective} transcription initiation rate $\alpha_{eff}$ {\bf (a)}, {\it effective} termination rate $\beta_{eff}$ {\bf (b)},  {\it bypass} transcription rate $\beta_{bp}$ {\bf (c)}, and the transcription {\it effectiveness} $r$ {\bf (d)}, as functions of the detachment rate $k_d$. The lines with marker \lq$\circ$' are for the cases with zero backtracking rate $k_b$, and the thick solid lines are calculated with large transcription termination rate $\beta$. In all calculations, the detachment rate $k_d$ changes from 0 to 1 with an increment 0.1. For other parameter values used in calculations, see Table \ref{paratable}. }\label{kddense}
\end{figure}

The plots in Fig. \ref{kddense}{\bf (a)} show that, generally, the {\it effective} initiation rate $\alpha_{eff}$ increases with the detachment rate of stalled polymerase at damaged site $l$. Because for large values of detachment rate,
the polymerase density between site 0 and site $l$ will be low. Therefore, the {\it effective} initiation rate $\alpha_{eff}=\alpha(1-p_0)$ will be large. But for the cases with large termination rate $\beta$, $\alpha_{eff}$ is almost independent of detachment rate $k_d$, see the thick solid line in Fig. \ref{kddense}{\bf (a)}. This is because, for large termination rate $\beta$, the polymerase density along transcription template is low enough, and the influence of detachment of stalled polymerase can be neglected. In other words, detachment will not help to reduce the polymerase density any longer. From the plots in Fig. \ref{kddense}{\bf (b)}, one can see that for low termination rate $\beta$, the {\it effective} transcription rate $\beta_{eff}$ increases with detachment rate $k_d$. The reason is that large detachment rate $k_d$ will be helpful to reduce the polymerase density along the transcript template. Consequently, the mean translocation speed of polymerase will be high. But for large values of termination rate $\beta$, $\beta_{eff}$ decreases with detachment rate $k_d$ (see the thick solid line in Fig. \ref{kddense}{\bf (b)}). Given large values of $\beta$, the polymerase density along template will be low enough such that each polymerase can translocate forward freely. Thus, with large values of detachment rate $k_d$, polymerase will have less opportunity to complete its whole transcription process. This means that the {\it effective} transcription rate $\beta_{eff}$ will be low for large detachment rate $k_d$. Because there are altogether three possible mechanisms for stalled polymerase to leave the damaged site $l$, i.e., backtracking, detachment, and bypass, the {\it bypass} transcription rate $\beta_{bp}$ will be low for large detachment rate $k_d$, see Fig. \ref{kddense}{\bf (c)}. The plots in Fig. \ref{kddense}{\bf (d)} show that, for the cases with nonzero backtracking rate $k_b$, transcription {\it effectiveness} $r$ decreases slightly with detachment rate $k_d$. But for the cases with zero backtracking rate, {\it effectiveness} $r$ increases with $k_d$. This implies that when no backtracking, detachment is one good mechanism to solve the polymerase stalling problem. But generally, backtracking may be better than detachment at increasing the transcription efficiency.

\begin{figure}
  \centering
  \includegraphics[width=12cm]{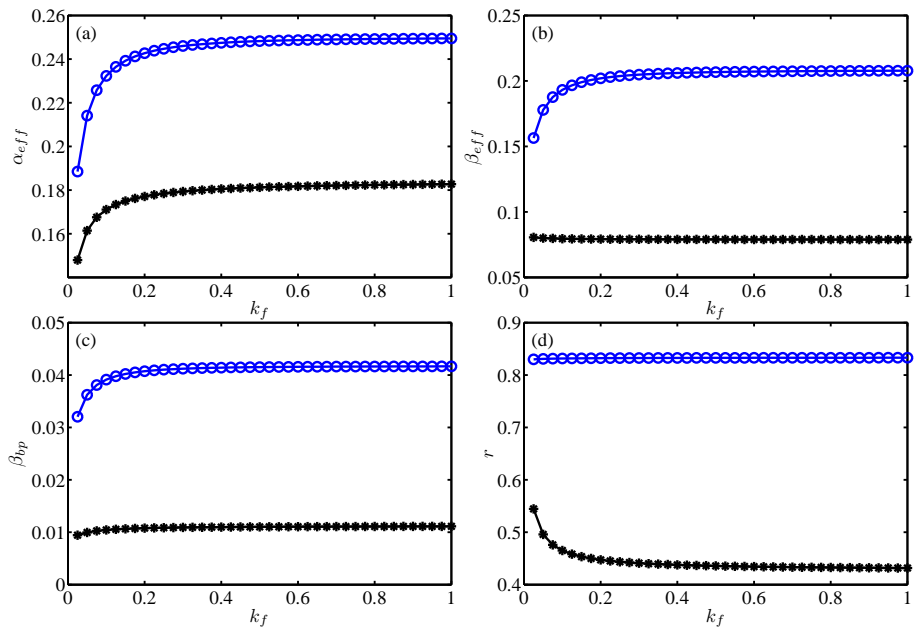}\\
  \caption{The {\it effective} transcription initiation rate $\alpha_{eff}$ {\bf (a)}, {\it effective} termination rate $\beta_{eff}$ {\bf (b)},  {\it bypass} termination rate $\beta_{bp}$ {\bf (c)}, and the transcription {\it effectiveness} $r$ {\bf (d)}, as functions of the forward rate $k_f$ (see Fig. \ref{model}{\bf (a)} for the meaning of $k_f$). The lines with marker \lq$\circ$' are obtained with zero detachment rate $k_d$ and high termination rate $\beta$, while the lines with marker \lq$\ast$' are obtained with nonzero detachment rate $k_d$ and low termination rate $\beta$. In all calculations, the forward rate $k_f$ changes from 0.1 to 1 with an increment 0.1. The values of other parameters are listed in Table \ref{paratable}. }\label{kfdense}
\end{figure}

Figs. \ref{kfdense}{\bf (a-c)} show that all the {\it effective} rates, $\alpha_{eff}$, $\beta_{eff}$, and $\beta_{bp}$ increase with the return back rate $k_f$ of the backtracked polymerase. Since large value of rate $k_f$ means that the backtracked polymerase at site $l-1$ will return back to site $l$ quickly when the damaged site $l$ has been repaired, and then restart its transcription. But the plots in Fig. \ref{kfdense}{\bf (d)} show that for nonzero detachment rate $k_d$ and low termination rate $\beta$, transcription {\it effectiveness} decreases with rate $k_f$. Given low termination rate $\beta$, the polymerase density between sites $l$ and $N$ will be high, so the increase of return back rate $k_f$ has little influence to increase the {\it effective} transcription rate $\beta_{eff}$ (see the line in Fig. \ref{kfdense}{\bf (b)} with marker \lq\lq$\ast$"). But for nonzero detachment rate $k_d$, the polymerase translocation between sites 0 and $l$ may be uncrowded, thus, the {\it effective} transcription initiation rate $\alpha_{eff}$ increases with the return back rate $k_f$ (see the line in Fig. \ref{kfdense}{\bf (a)} with marker \lq\lq$\ast$"). Therefore, from the definition of transcription {\it effectiveness}, $r:=\beta_{eff}/\alpha_{eff}$, for the cases with low termination rate $\beta$ but nonzero detachment rate $k_d$,  transcription {\it effectiveness} $r$ decreases with the return back rate $k_f$. Therefore, large return back rate $k_f$ may not be helpful to increase the efficiency of transcription.

\begin{figure}
  \centering
  \includegraphics[width=12cm]{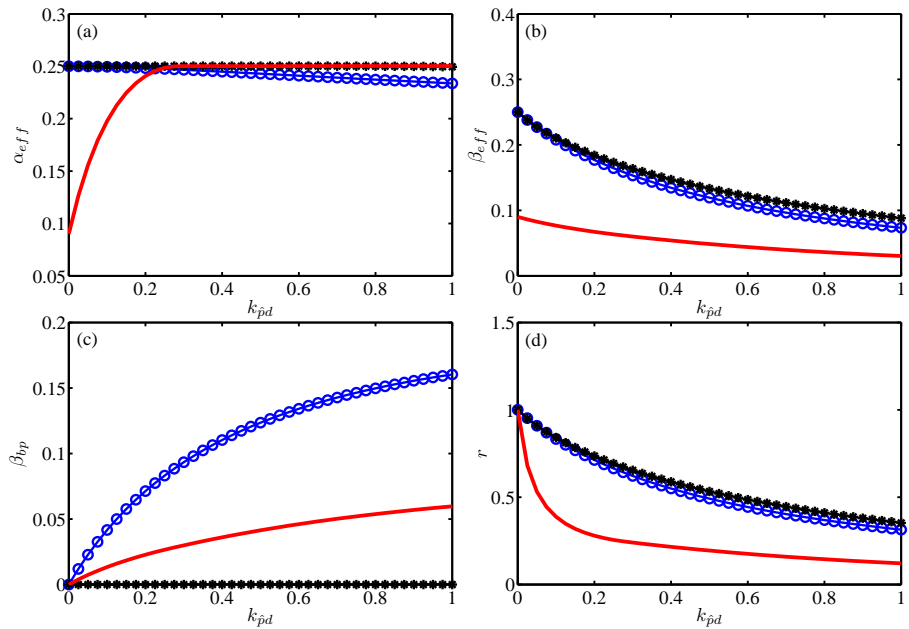}\\
  \caption{The {\it effective} transcription initiation rate $\alpha_{eff}$ {\bf (a)}, {\it effective} termination rate $\beta_{eff}$ {\bf (b)},  {\it bypass} termination rate $\beta_{bp}$ {\bf (c)}, and the transcription {\it effectiveness} $r$ {\bf (d)}, as functions of the damage rate  $k_{\hat{p}d}$. The lines with marker \lq$\circ$' are obtained with zero detachment rate $k_d$, the lines with marker \lq$\ast$' are obtained with zero bypass rate $k_{bp}$, and the thick solid lines are obtained with zero backtracking rate $k_b$ and low termination rate $\beta$. In all calculations, the damage rate $k_{\hat{p}d}$ changes from 0 to 1 with an increment 0.1. The values of other parameters are listed in Table \ref{paratable}.}\label{khatpddense}
\end{figure}

Finally, the plots in Figs. \ref{khatpddense}{\bf (b-d)} show that the {\it effective} transcription rate $\beta_{eff}$ and transcription {\it effectiveness} $r$ decrease with the damage rate $k_{\hat{p}d}$ of site $l$, while the {\it bypass} transcription rate $\beta_{bp}$ increases with $k_{\hat{p}d}$. For high damage rate $k_{\hat{p}d}$, the polymerase is more likely to be stalled at the site $l$, and then the possibility of bypass will be high and the synthesis speed of correct mRNA will be low. The plots in Fig. \ref{khatpddense}{\bf (a)} imply that, for low termination rate $\beta$ and low backtracking rate $k_b$, the {\it effective} initiation rate $\alpha_{eff}$ increases with damage rate $k_{\hat{p}d}$. Given low termination rate $\beta$, the polymerase density along transcription template will be high and the translocation speed of polymerase will be low. With the increase of  damage rate $k_{\hat{p}d}$, the stalled polymerase will have more possibility to detach from the template. So the total leaving rate of polymerase from the transcription template, either from the stop site $N$ or from the damaged site $l$, will increase. Thus, the {\it effective} initiation rate $\alpha_{eff}$ increases with the damage rate $k_{\hat{p}d}$. The line with marker \lq\lq$\circ$" in Fig. \ref{khatpddense}{\bf (a)} also show that, without detachment, the {\it effective} initiation rate $\alpha_{eff}$ decreases with damage rate $k_{\hat{p}d}$. This is because, for large damage rate, polymerase will be more likely to be stalled at the damaged site $l$, and consequently, the translocation speed of polymerase along transcription template will be slowed down.

\section{Concluding Remarks}
In this study, a modified TASEP model is presented to describe gene transcription process including polymerase stalling. Because of the detachment (or degradation) of polymerase from the damaged site, the polymerase density along transcription template has a sharp change at the damaged site. As in usual TASEP models, the polymerase density may have boundary layers at the transcription start site and termination site. In the main body of the transcription template, the polymerase density may be in three phases, low density phase, high density phase, and maximal flux phase. But the phases in different regions of the transcription template may be different. This study showed that the {\it effective} transcription rate (the rate to synthesize mRNA correctly), and the transcription {\it effectiveness} (the proportion of correct transcription) will be high if the damaged site of the transcription template is close to the transcription termination site. The increase of transcription initiation rate will increase the {\it effective} transcription rate, but decrease the transcription {\it effectiveness}. On the other hand, the increase of transcription termination rate will increase the {\it effective} transcription rate and the transcription {\it effectiveness}, as well as the {\it effective} transcription initiation rate.

Experiments found that there are three mechanisms for cells to solve the polymerase stalling problem, backtracking, bypass and detachment. This study showed that the increase of backtracking rate will lead to the increase of {\it effective} transcription rate and the transcription {\it effectiveness}, but lead to the decrease of the {\it effective} transcription initiation rate. It implies that backtracking is one of the ideal mechanisms to increase the synthesizing rate of mRNA and the transcription efficiency. Without backtracking and detachment, the increase of bypass rate will lead to the increase of {\it effective} transcription rate. But for general cases, large values of bypass rate will lead to low values of {\it effective} transcription rate and the transcription {\it effectiveness}. Similarly, without backtracking, detachment (or degradation) of the stalled polymerase is one good mechanism to solve the stalling problem. But for nonzero backtracking rate cases, the increase of detachment rate may lead to the decrease of {\it effective} transcription rate and transcription {\it effectiveness}. As expected, the increase of damage rate of the transcription template will lead to the decrease of transcription efficiency.

The results obtained in this study will be helpful to the understandings of gene transcription in living cells, and the mechanisms used by cells to solve the polymerase stalling problem. The model presented in this study can be further generalized to discuss more general cases of gene transcription process in which polymerase may be stalled at more than one site of the transcription template. The model parameter values in real cells may be extracted through the NET-seq approach presented in \cite{Churchman2011}, and then the theoretical model given in this study can be used to do quantitative analysis of the gene transcription process with polymerase stalling. Finally, combining this study with the recent model presented by Choubey {\it et al} in \cite{Choubey2014}, more details of the transcript process can be better understood.

\vskip 0.5cm

\begin{acknowledgements}
This study was supported by the Natural Science Foundation of China (Grant No. 11271083), and the National Basic Research Program of China (National \lq\lq973" program, project No. 2011CBA00804).
\end{acknowledgements}


\begin{thebibliography}{10}

\bibitem{Krug1991}
Joachim Krug.
\newblock Boundary-induced phase transitions in driven diffusive systems.
\newblock {\em Phys. Rev. Lett.}, 67:1882--1885, 1991.

\bibitem{Derrida1993}
B.~Derrida, S.~A. Janowsky, J.~L. Lebowitz, and E.~R. Speer.
\newblock Exact solution of the totally asymmetric simple exclusion process:
  Shock profiles.
\newblock {\em Journal of Statistical Physics}, 73:813--842, 1993.

\bibitem{Schutz1993}
G.~Schutz and E.~Domany.
\newblock Phase transitions in an exactly soluble one-dimensional exclusion
  process.
\newblock {\em Journal of Statistical Physics}, 72:277--296, 1993.

\bibitem{Kolomeisky1998}
A.~Kolomeisky, G.M. Schutz, E.B. Kolomeisky, and J.P. Straley.
\newblock Phase diagram of one-dimensional driven lattice gases with open
  boundaries.
\newblock {\em J. Phys. A: Math. Gen.}, 31:6911--6919, 1998.

\bibitem{Zhang20113}
Yunxin Zhang.
\newblock Periodic one-dimensional hopping model with transitions between
  nonadjacent states.
\newblock {\em Phys. Rev. E}, 84:031104, 2011.

\bibitem{Zhang20131}
Yunxin Zhang.
\newblock Theoretical analysis of kinesin {KIF1A} transport along microtubule.
\newblock {\em J. Stat. Phys.}, 152:1207--1221, 2013.

\bibitem{Azvolinsky2009}
Anna Azvolinsky, Paul~G. Giresi, Jason~D. Lieb, and Virginia~A. Zakian.
\newblock Highly transcribed {RNA} polymerase {II} genes are impediments to
  replication fork progression in saccharomyces.
\newblock {\em Molecular cell}, 34(6):722--734, June 2009.

\bibitem{Dalgaard2001}
Jacob~Z. Dalgaard and Amar~J.S. Klar.
\newblock A {DNA} replication-arrest site rts1 regulates imprinting by
  determining the direction of replication at mat1 in {S}. pombe.
\newblock {\em Genes $\&$ development}, 15(16):2060--2068, August 2001.

\bibitem{Brewer1988}
Bonita~J. Brewer and Walton~L. Fangman.
\newblock A replication fork barrier at the 3' end of yeast ribosomal {RNA}
  genes.
\newblock {\em Cell}, 55(4):637--643, November 1988.

\bibitem{Gruber2000}
Markus Gruber, Ralf~Erik Wellinger, and Jos${\rm \acute{e}}$~M. Sogo.
\newblock Architecture of the replication fork stalled at the 3' end of yeast
  ribosomal genes.
\newblock {\em Molecular and cellular biology}, 20(15):5777--5787, August 2000.

\bibitem{Nouspikel2002}
Thierry Nouspikel and Philip~C. Hanawalt.
\newblock {DNA} repair in terminally differentiated cells.
\newblock {\em DNA repair}, 1(1):59--75, January 2002.

\bibitem{Casper2002}
Anne~M. Casper, Paul Nghiem, Martin~F. Arlt, and Thomas~W. Glover.
\newblock {ATR} regulates fragile site stability.
\newblock {\em Cell}, 111(6):779--789, December 2002.

\bibitem{Cha2002}
Rita~S. Cha and Nancy Kleckner.
\newblock {ATR} homolog {M}ec1 promotes fork progression, thus averting breaks
  in replication slow zones.
\newblock {\em Science}, 297(5581):602--606, July 2002.

\bibitem{Andreea2014}
Daraba Andreea, Gali~Vamsi K, Halmai Mikl${\rm \acute{o}}$s, Haracska Lajos,
  and Unk Ildiko.
\newblock {D}ef1 promotes the degradation of {P}ol3 for polymerase exchange to
  occur during {DNA}-damage-induced mutagenesis in saccharomyces cerevisiae.
\newblock {\em PLoS Biology}, 12(1):e1001771, January 2014.

\bibitem{Damsma2007}
Gerke~E Damsma, Aaron Alt, Florian Brueckner, Thomas Carell, and Patrick
  Cramer.
\newblock Mechanism of transcriptional stalling at cisplatin-damaged {DNA}.
\newblock {\em Nature structural $\&$ molecular biology}, 14(12):1127--1133,
  December 2007.

\bibitem{Walmacq2012}
Celine Walmacq, Alan~C.M. Cheung, Maria~L. Kireeva, Lucyna Lubkowska,
  Chengcheng Ye, Deanna Gotte, Jeffrey~N. Strathern, Thomas Carell, Patrick
  Cramer, and Mikhail Kashlev.
\newblock Mechanism of translesion transcription by {RNA} polymerase {II} and
  its role in cellular resistance to {DNA} damage.
\newblock {\em Molecular cell}, 46(1):18--29, April 2012.

\bibitem{Edenberg2014}
Ellen~R. Edenberg, Michael Downey, and David Toczyski.
\newblock Polymerase stalling during replication, transcription and
  translation.
\newblock {\em Current Biology}, 24(10):R445--R452, May 2014.

\bibitem{Giannattasio2004}
Michele Giannattasio, Federico Lazzaro, Maria~Pia Longhese, Paolo Plevani, and
  Marco Muzi-Falconi.
\newblock Physical and functional interactions between nucleotide excision
  repair and {DNA} damage checkpoint.
\newblock {\em EMBO journal}, 23(2):429--438, January 2004.

\bibitem{Ciccia2012}
Alberto Ciccia, Amitabh~V. Nimonkar, Yiduo Hu, Ildiko Hajdu, Yathish~Jagadheesh
  Achar, Lior Izhar, Sarah~A. Petit, Britt Adamson, John~C. Yoon, Stephen~C.
  Kowalczykowski, David~M. Livingston, Lajos Haracska, and Stephen~J. Elledge.
\newblock Polyubiquitinated {PCNA} recruits the {ZRANB}3 translocase to
  maintain genomic integrity after replication.
\newblock {\em Molecular cell}, 47(3):396--409, August 2012.

\bibitem{Weston2012}
Ria Weston, Hanneke Peeters, and Dragana Ahel.
\newblock {ZRANB}3 is a structure-specific {ATP}-dependent endonuclease
  involved in replication stress response.
\newblock {\em Genes $\&$ development}, 26(14):1558--1572, July 2012.

\bibitem{Yuan2012}
Jingsong Yuan, Gargi Ghosal, and Junjie Chen.
\newblock The {HARP}-like domain-containing protein {AH}2/{ZRANB}3 binds to
  {PCNA} and participates in cellular response to replication stress.
\newblock {\em Molecular cell}, 47(3):410--421, August 2012.

\bibitem{Vermeulen2013}
Wim Vermeulen and Maria Fousteri.
\newblock Mammalian transcription-coupled excision repair.
\newblock {\em Cold Spring Harbor perspectives in biology}, 5(8):a012625,
  August 2013.

\bibitem{Lagerwerf2011}
Saskia Lagerwerf, Mischa~G. Vrouwe, Ren${\rm \acute{e}}$~M. Overmeer, Maria~I.
  Fousteri, and Leon~H.F. Mullenders.
\newblock {DNA} damage response and transcription.
\newblock {\em DNA Repair}, 10(7):743--750, July 2011.

\bibitem{Deaconescu2013}
Alexandra~M. Deaconescu.
\newblock {RNA} polymerase between lesion bypass and {DNA} repair.
\newblock {\em Cellular and molecular life sciences}, 70(23):4495--4509,
  December 2013.

\bibitem{Higgins1976}
N.P. Higgins, K.~Kato, and B.~Strauss.
\newblock A model for replication repair in mammalian cells.
\newblock {\em Journal of Molecular Biology}, 101(3):417--425, March 1976.

\bibitem{Stelter2003}
Philipp Stelter and Helle~D. Ulrich.
\newblock Control of spontaneous and damage-induced mutagenesis by {SUMO} and
  ubiquitin conjugation.
\newblock {\em Nature}, 425(6954):188--191, September 2003.

\bibitem{Watanabe2004}
Kenji Watanabe, Satoshi Tateishi, Michio Kawasuji, Toshiki Tsurimoto, Hirokazu
  Inoue, and Masaru Yamaizumi.
\newblock Rad18 guides pol$\eta$ to replication stalling sites through physical
  interaction and {PCNA} monoubiquitination.
\newblock {\em EMBO journal}, 23(19):3886--3896, September 2004.

\bibitem{Wilson2013}
Marcus~D. Wilson, Michelle Harreman, and Jesper~Q. Svejstrup.
\newblock Ubiquitylation and degradation of elongating {RNA} polymerase {II}:
  The last resort.
\newblock {\em Biochimica et Biophysica Acta}, 1829(1):151--157, January 2013.

\bibitem{Shoemaker2012}
Christopher~J. Shoemaker and Rachel Green.
\newblock Translation drives m{RNA} quality control.
\newblock {\em Nature structural $\&$ molecular biology}, 19(6):594--601, June
  2012.

\bibitem{Doma2006}
Meenakshi~K. Doma and Roy Parker.
\newblock Endonucleolytic cleavage of eukaryotic m{RNA}s with stalls in
  translation elongation.
\newblock {\em Nature}, 440(7083):561--564, March 2006.

\bibitem{Becker2011}
Thomas Becker, Jean-Paul Armache, Alexander Jarasch, Andreas~M Anger, Elizabeth
  Villa, Heidemarie Sieber, Basma~Abdel Motaal, Thorsten Mielke, Otto
  Berninghausen, and Roland Beckmann.
\newblock Structure of the no-go m{RNA} decay complex {D}om34-{H}bs1 bound to a
  stalled 80{S} ribosome.
\newblock {\em Nature structural $\&$ molecular biology}, 18(6):715--720, June
  2011.

\bibitem{Shoemaker2010}
Christopher~J. Shoemaker, Daniel~E. Eyler, and Rachel Green.
\newblock {D}om34:{H}bs1 promotes subunit dissociation and peptidyl-t{RNA}
  drop-off to initiate no-go decay.
\newblock {\em Science}, 330(6002):369--372, October 2010.

\bibitem{Li2014}
Jingwei Li and Yunxin Zhang.
\newblock Relationship between promoter sequence and its strength in gene
  expression.
\newblock {\em Eur. Phys. J. E}, 37:86, 2014.

\bibitem{Derrida19931}
B.~Derrida, M.R. Evans, V.~Hakim, and V.~Pasquier.
\newblock Exact solution of a 1d asymmetric exclusion model using a matrix
  formulation.
\newblock {\em J. Phys A: Math. Gen.}, 26:1493--1517, 1993.

\bibitem{Churchman2011}
L.~Stirling Churchman and Jonathan~S. Weissman.
\newblock Nascent transcript sequencing visualizes transcription at nucleotide
  resolution.
\newblock {\em Nature}, 469:368--373, 2011.

\bibitem{Choubey2014}
Sandeep Choubey, Alvaro Sanchez, and Jane Kondev.
\newblock Deciphering transcriptional dynamics in vivo by counting nascent rna
  molecules.
\newblock {\em arXiv:1311.0050v2}, 2014.

\end{thebibliography}
\end{document}